\documentclass{article}

\def\copyrighttext{%
\renewcommand{\baselinestretch}{0.8}
  \small Copyright 2019 IEEE. Published in the 2019 IEEE Automatic Speech
  Recognition and Understanding Workshop (ASRU 2019), scheduled for
  December 14-18, 2019 in Sentosa, Singapore.
  \footnotesize  Personal use of this material is permitted.
  However, permission to reprint/republish this material for advertising or
  promotional purposes or for creating new collective works for resale or
  redistribution to servers or lists, or to reuse any copyrighted component of
  this work in other works, must be obtained from the IEEE. Contact: Manager,
  Copyrights and Permissions / IEEE Service Center / 445 Hoes Lane / P.O. Box
  1331 / Piscataway, NJ 08855-1331, USA. Telephone: + Intl. 908-562-3966.}

\usepackage{spconf,amsmath,graphicx}
\usepackage{booktabs}
\usepackage{units}
\usepackage{xspace}
\usepackage{refcount}

\ifdefined\copyrighttext
\usepackage{tikz}
\newcommand\ieeecopyrightnotice{%
\begin{tikzpicture}[remember picture,overlay]
\node[anchor=south,yshift=8pt] at (current page.south) {\fbox{\parbox{\dimexpr\textwidth-\fboxsep-\fboxrule\relax}{\copyrighttext}}};
\end{tikzpicture}%
}
\else
\newcommand\ieeecopyrightnotice{\relax}
\fi

\newcommand{\citep}{\cite}  
\newcommand{\citet}{\cite}  
\newcommand{\ytdev}{\textsc{YTDev18}\xspace}
\newcommand{\lrsted}{\textsc{LRS3-TED}\xspace}
\newcommand{\negsectionspace}{\vspace{-4pt}}
\newcommand{\negfigspace}{\vspace{-4pt}}

\title{Recurrent Neural Network Transducer for Audio-Visual Speech
  Recognition}

\ifdefined\blindreview
\name{[BLIND]}
\address{[BLIND]}
\else
\name{
\parbox{\textwidth}{
\centering
Takaki Makino$^{\dag 1}$ \quad
Hank Liao$^{\dag 2}$ \quad
Yannis Assael$^\ddag$ \quad
Brendan Shillingford$^\ddag$ \\
Basilio Garcia$^\dag$ \quad
Otavio Braga$^\dag$ \quad
Olivier Siohan$^\dag$ \\}
}
\address{
$^\dag$ Google Inc., 1600 Amphitheatre Pkwy, Mountain View, CA 94043, USA \\
$^\ddag$ DeepMind, 6 Pancras Square, Kings Cross, London N1C 4AG, UK \\
$^1$\texttt{tmakino@google.com}, $^2$\texttt{hankliao@google.com}
}
\fi

\begin{document}

\maketitle

\begin{abstract}
  This work presents a large-scale audio-visual speech recognition
  system based on a recurrent neural network transducer (RNN-T)
  architecture. To support the development of such a system, we built
  a large audio-visual (A/V) dataset of segmented utterances extracted
  from YouTube public videos, leading to 31k hours of audio-visual
  training content. The performance of an audio-only, visual-only, and
  audio-visual system are compared on two large-vocabulary test sets:
  a set of utterance segments from public YouTube videos called \ytdev
  and the publicly available \lrsted set. To highlight the
  contribution of the visual modality, we also evaluated the
  performance of our system on the \ytdev set artificially corrupted
  with background noise and overlapping speech. To the best
  of our knowledge, our system significantly improves the
  state-of-the-art on the \lrsted set.
\ieeecopyrightnotice
\ifnum \getpagerefnumber{endofmaintext} > 6
\smash{\begin{picture}(0,0)
   \put(0,0){\Huge DO NOT SUBMIT, \getpagerefnumber{endofmaintext} PAGES USED}
\end{picture}}
\fi
\end{abstract}


\negsectionspace
\section{Introduction}
\negsectionspace

While the performance of automatic speech recognition (ASR) systems
has significantly improved over the past several years, outstanding
challenges remain for ubiquitous ASR. In particular, state-of-the-art
recognizers can fail in noisy environments or in presence of
overlapping speech. In some applications such as auto-captioning of
videos, multiple modalities are available when transcribing speech. The
visual signal can supply the spelling of obscure names and
terms when shown on-screen, or provide contextual information related
to the visual scene~\citet{gupta2017}. Further, the motion of the lips
constrains the possible phonemes and hence words that can be
spoken. With the availability of sophisticated neural network
architectures and large amounts of multimedia data, it is intriguing
to explore how audio and visual modalities can be combined to yield
improvements not otherwise available in a unimodal setting.

Our work is motivated by the recent success in automatic lip reading
by~\citet{dm_lsvsr_2018} and audio-visual speech recognition
in~\citet{ox_davsr_2018,Petridis2018AudioVisualSR}. In the former, a
state-of-the-art Vision-to-Phoneme model (V2P) is developed using a
carefully and unprecedentedly-sized visual corpus of speaking faces
with their corresponding text transcripts extracted from YouTube
public videos. A large VGG-inspired neural network was used for
extracting visual features, with a connectionist temporal
classification (CTC) based system~\citet{ctc06} to predict
phonemes. In the latter, using a large audio-visual data set with
captions derived mainly from British television, the authors were able
to use transformer-based self-attention network
blocks~\citet{xformer17nips} to model the audio and visual modalities
with a sequence-to-sequence loss~\citet{seq2seq14cho}. With their
approach called TM-seq2seq, they demonstrate that combining audio and
visual features yields better performance than audio-only models on a
large vocabulary recognition task of transcribing TED talk videos,
released as a data set called \lrsted~\citet{lrs3ted}.

As for most machine learning tasks, the performance of an AV-ASR
system is driven by the availability of high-quality training
datasets. In this work, we first use the approach described
in~\citet{liao13asru} to mine a 150k hours, audio-only speech
recognition corpus from YouTube with labels derived from user-uploaded
captions. Next, we filter this dataset using face tracking technology
similar to the work in~\citet{lrs3ted,dm_lsvsr_2019} to select
segmented utterances with a matching on-screen speaking face, leading
to a 31k hours audio-visual corpus.
Similar to~\citet{dm_lsvsr_2019}, we apply face landmark smoothing to
stabilize the face thumbnails. But in contrast to previous AV-ASR
work, we synchronize audio to video frames by adjusting the audio
feature frame rate while maintaining a fixed short-term Fourier
transform analysis window.

In terms of modeling, as observed
in~\citet{ox_davsr_2018,Afouras2018DeepLR}, a weakness of CTC modeling
as used in V2P is that the network outputs are conditionally
independent of each other and require an external language model, such
as a N-gram Finite State Transducer to perform well. In contrast,
using a sequence-to-sequence loss allows the language model to be
learned implicitly as part of the decoder. We advocate using the
recurrent neural network transducer (RNN-T), first described
in~\citep{rnnt12graves}, as a more natural model for speech
recognition. The RNN-T consists of two components, a transcription or
encoder network and a recurrent prediction or decoder network. The
encoder can be viewed as the acoustic or visual model, with in our
case, inputs being mel-filterbank coefficients~\citet{Davis80} or
visual embeddings derived using a VGG inspired
network~\citet{vgg15simonyan} as 3-D convolutions in V2P. The decoder
network models the output of the current symbol given the previous
output symbol, in our case character-level outputs. The encoder and
decoder outputs are fed into a joint network where the output is
normalized to yield a softmax over characters. We believe that the
RNN-T is a better model for speech since the output dependencies tend
to be local and the use of LSTM layers for the encoder and decoder
better models this locality. Compared to sequence-to-sequence models,
the use of attention may be less desirable since the model spends
capacity learning that the most important features for the output are
the nearest input frames. In other words, for speech recognition the
alignment of input features to output symbols is monotonic. The use of
character level outputs, rather than phonemes, provides a simple way
of achieving an open-vocabulary system.

We evaluate our approach on a general YouTube transcription task for
audio, video, and audio-visual speech recognition. To illustrate the
impact of the visual modality on ASR performance, we artificially
corrupted our evaluation dataset by adding either babble noise at
various signal-to-noise ratios (SNR), or a few seconds of overlapping
speech at the beginning or end of each evaluation utterance. We also
provide experimental results on the publicly available
large-vocabulary benchmark \lrsted~\citet{lrs3ted} and report, to the
best of our knowledge, state-of-the-art results on that set.

The paper is organized as follows. In Section~\ref{Datasets} we
describe the way we construct our training set and present our
test sets. Section~\ref{System Architecture} focuses on the
architecture of our system, starting with our process to generate
synchronized audio and visual features, followed by our RNN-T approach
that operates on either audio, visual, or AV features. Experiments
are discussed in Section~\ref{Experiments}, Section~\ref{principles}
highlights the AI Principles followed in this work, and
Section~\ref{Conclusions} concludes the paper.


\negsectionspace
\section{Datasets}
\label{Datasets}
\negsectionspace
The impact of the visual modality on human speech perception has been
documented as early as 1954~\citet{sumby54}. In particular, the place
of articulation can provide valuable information to help differentiate
acoustically confusable sounds~\citet{massaro98}. As a result, and as
highlighted in several reviews of AV-ASR
technology~\citet{Potamianos03,Czyzewski2017}, attempts were made in
the early 90's to combine audio and visual modalities for ASR. Until a
few years ago, the large majority of the publicly available
AV datasets suitable for AV-ASR were limited in size and scope,
consisting mostly of tens of hours of spoken digits or
command-and-control word sequences with a limited number of speakers
recorded in a laboratory environment under controlled lighting
conditions and face orientation~\citet{Czyzewski2017}.

In 2016, researchers proposed a novel end-to-end trained model for
visual speech recognition~\citet{assael2016lipnet} and approach to
construct large AV datasets suitable for
AV-ASR~\citet{Chung16}. Starting from TV shows and their corresponding
closed-captions provided by the BBC, they applied computer vision
techniques to select short audio segments with an on-screen speaking
face matching the audio. In subsequent
studies~\citet{Chung17,Chung17a}, they extended their approach to a
larger amount of data, leading to the LRS2 \& LRS3 datasets consisting
of videos from the BBC and TED-talks, respectively, for a total of 800
hours of transcribed AV content.  More recently, applying this method
to public videos from YouTube lead to an AV-ASR dataset consisting of
3.9k hours of videos~\citet{dm_lsvsr_2018}.

In this paper, we expand the approach of~\citet{dm_lsvsr_2018} to
construct a significantly larger dataset. As in~\citet{dm_lsvsr_2018},
we build on the semi-supervised approach described
in~\cite{liao13asru} to mine a large audio dataset from YouTube
videos. When YouTube users and content creators upload a public video
on YouTube, they have the option of uploading subtitles alongside
their video. A forced alignment procedure is then used to synchronize
the user-uploaded transcripts with the audio to generate time-aligned
captions. We take advantage of user-uploaded captions to automatically
mine our AV dataset. First, we run ASR on the uploaded video and
string-align the ASR transcripts with the user-uploaded
transcripts. Audio segments where the ASR and user-uploaded
transcripts agree are selected as training utterances, treating the
user-uploaded transcripts as ground-truth reference. Because of the
scale of YouTube, this leads to the construction of a large audio
training set, in the order of 150k hours of data for American
English. Next, we extract the video snippets corresponding to the
selected utterances and run face tracking to locate all on-screen
faces~\citet{cloud_api,facenet15schroff}. A visual speech classifier
is used to identify the speaking face, if any, that spans each audio
segment, followed by an audio-visual synchrony detector to reject
dubbed videos.

The result of that process is a collection of short utterances (from a
few seconds to tens of seconds long) totaling 31k hours of data, where
with high confidence, the audio matches the user-uploaded transcript,
and the selected face video track matches the audio.  This exceeds the
amount of training data used for training the V2P model
of~\citet{dm_lsvsr_2018} by a factor of 5, and the TM-seq2seq model
of~\citet{seq2seq14cho} by a factor of 10. The size of our training
set is mitigated by the fact that we do not augment the training data
by perturbing the audio or image data. We believe the main reason we
obtain much more data than in the V2P work is due to accepting a
greater range of face pan and tilt angles, i.e. more than +/- 30
degrees. Unlike the LRW/LRS2/LRS3 datasets that are restricted to
professionally generated video content, our dataset spans a much
greater variety of content of speaking faces in the wild. We use this
dataset to train our unimodal audio and video systems, as well as our
AV ASR system, and extract ~70 hours of data as development set to
tune our models.

Unlike our training set that is automatically constructed from videos
with user-uploaded captions, we built our evaluation set from a
collection of manually transcribed YouTube videos. Starting from a set
of 1000 hours of transcribed videos, we applied the same face tracking
process as in training to select a collection of utterances with a
matching on-screen speaking face. Out of that process, we retained a
set of 20k segmented utterances, with their corresponding manual
reference transcripts and face tracks, called \ytdev and totaling 25
hours of data.
We checked the video IDs to confirm that the videos used in
the \ytdev and LRS3-TED evaluation sets are not included
in the training data.

To study the impact of the visual modality on recognition performance,
we artificially corrupt the \ytdev utterances in two ways. First, we
add babble noise randomly selected from the NoiseX~\citet{noisex}
dataset at 0dB, 10dB, and 20dB SNR to each utterance. Second, we add
overlapping speech from a competing speaker at the beginning/end of
each utterance at an equal energy level.

\negsectionspace
\section{System Architecture}
\label{System Architecture}
\negsectionspace

\subsection{Synchronized Audio-Visual Frames}
\label{sec:syncav}
\negsectionspace
A challenge in audio-visual speech recognition is dealing with the
difference in data rates between audio and visual features. In this
work we start with audio sampled and mixed down to a 16kHz,
single-channel signal using ffmpeg. A 25ms Hanning window with 10 ms
shift is used to compute a short-time Fourier transform for a spectral
representation of the audio. A mel-spaced bank of 80 triangular
filters is applied followed by a log function to yield mel filterbank
coefficients~\citet{Davis80}. These are stacked 2 frames to the left
and 2 to the right to yield a 400 dimensional feature vector, while
only keeping one in three frames similar to~\citet{ctcph15senior}, for
a final frame rate of 33$\nicefrac{1}{3}$ fps.

Because our data is heterogeneous in terms of video standards and frame
rates, we follow an approach similar to~\citet{dm_lsvsr_2018} and
downsample high frame rates down to a maximum of 30 fps, retaining
only videos between 23 fps to 30 fps, so that 23.98/24 fps (cinemas),
25 fps (PAL/SECAM), 29.97 fps (NTSC) and 30 fps videos are kept.
FaceNet~\citet{facenet15schroff} is applied to detect and track
faces. Face landmarks are then extracted from these tracks and
smoothed using a temporal Gaussian kernel. A difference from the V2P
data processing pipeline is that we do not filter out segments with
extreme face pan or tilt angles to train a more robust model, however
we do enforce a minimum eye distance of 80 pixels. Crops around the
mouth are then extracted as 128x128 RGB pixel thumbnails.

\begin{figure}[t]
\negfigspace
  \centering
  \includegraphics[width=\linewidth]{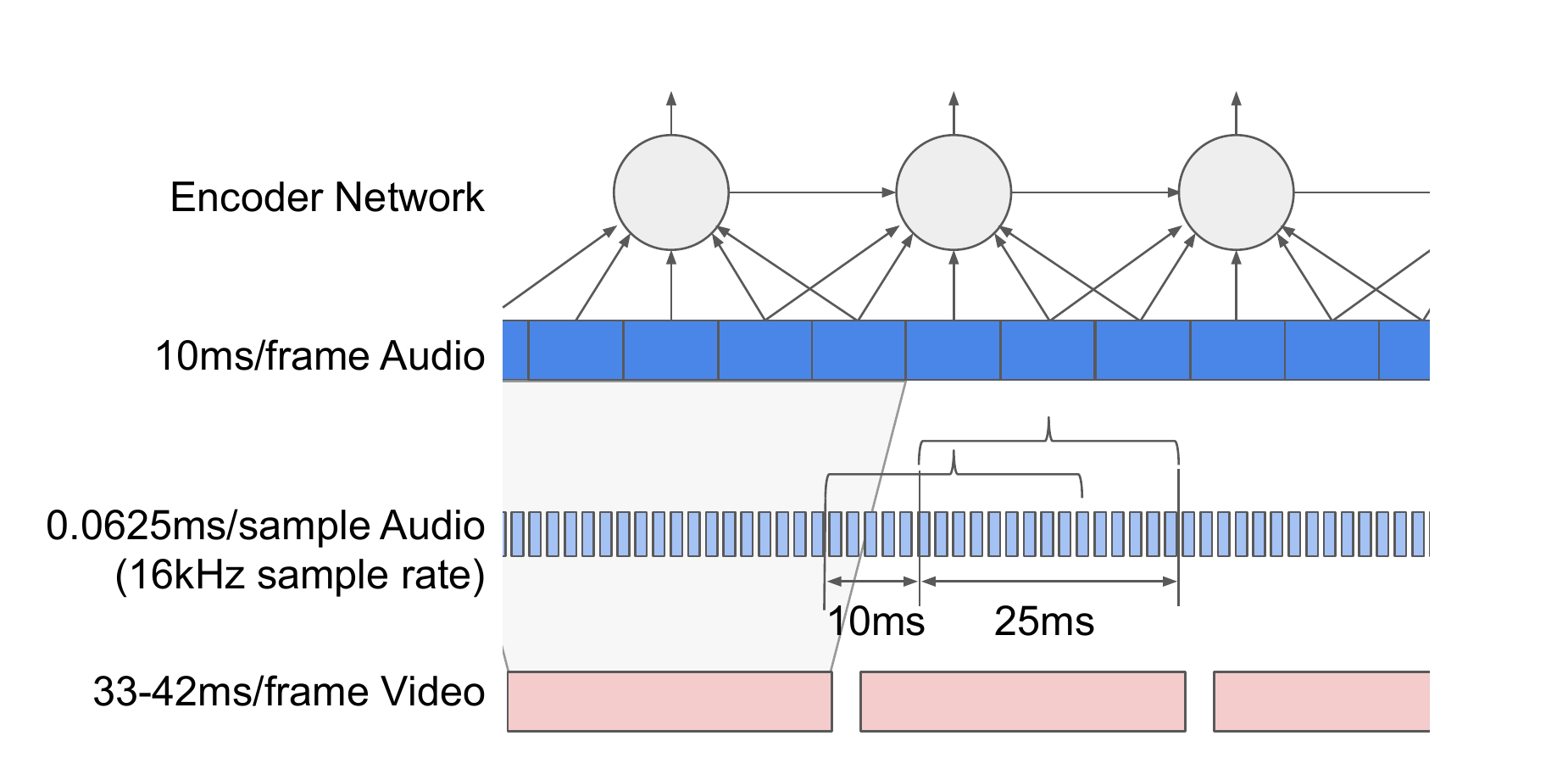}
  \caption{Unsynchronized audio and visual frames. Stacked mel-spaced
    filterbank features occur at a 30ms frame rate.  Video thumbnails
    occur at a 33-40 ms frame rate (25-30 frames per second).}
  \label{fg:unsyncfeat}

  \includegraphics[width=\linewidth]{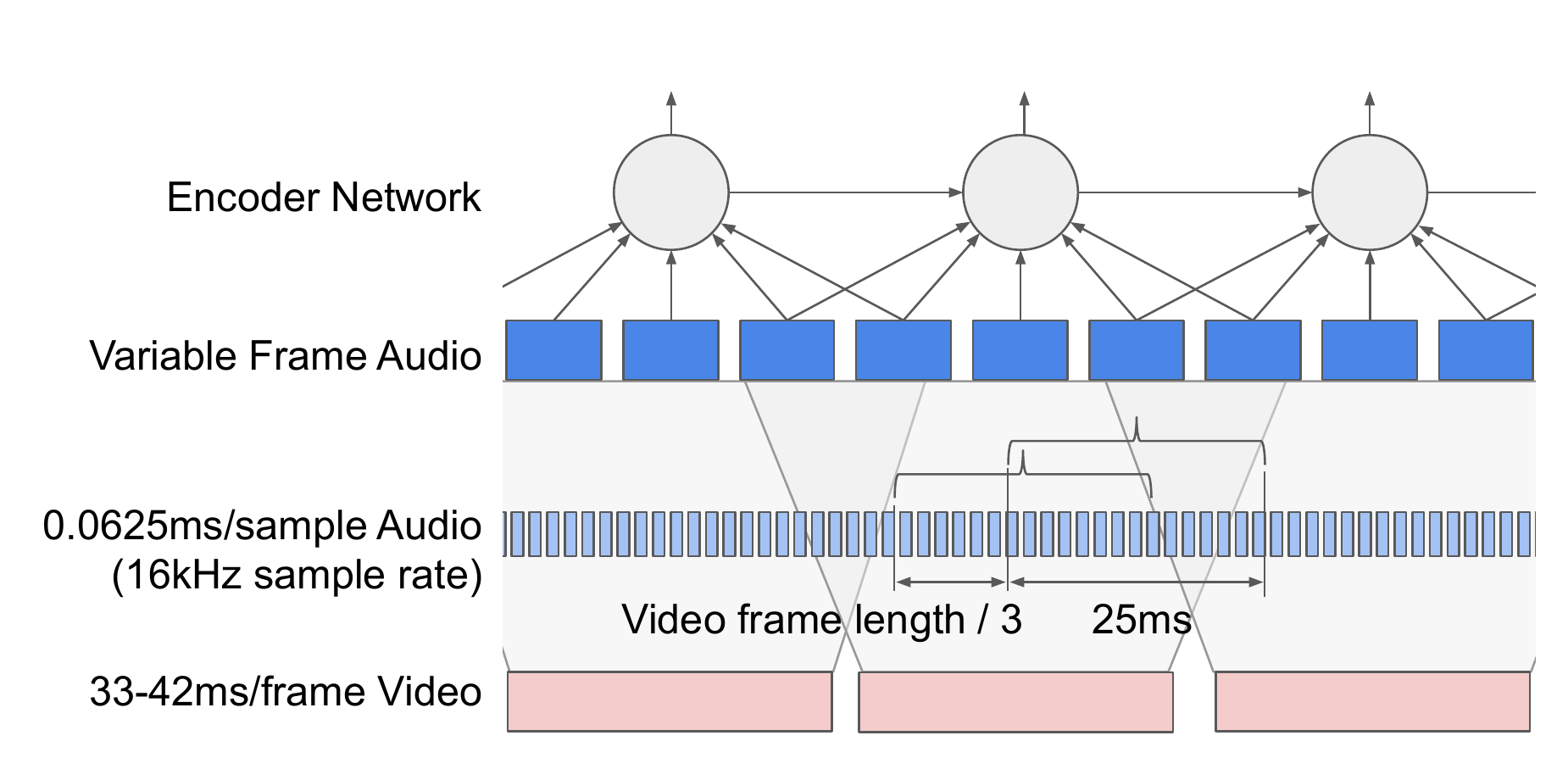}
  \caption{Synchronized audio and visual frames. Stacked mel-spaced
    filterbank features occur at a variable frame rate, matching video
    rate. Video thumbnails occur at a 33-42 ms frame rate (30-24
    frames per second). While the audio STFT analysis window remains
    25ms, the shift is variable.}
  \label{fg:syncfeat}
\negfigspace
\vspace{-6pt}
\end{figure}

Variable video frame rates typically do not match the audio
frame rate leading to unsynchronized audio and visual features as
illustrated in Figure~\ref{fg:unsyncfeat}. In~\citet{ox_davsr_2018},
the video is down-sampled to 25 fps and the audio feature stacked and
sampled to match this frame rate to yield synchronized frame rates. We
believe that the down-sampling of the video stream may degrade the
visual features and instead propose to operate on the
variable video frame rate and extract audio features at a matching
rate. Since the original audio has a high sampling rate, the 25 ms
STFT window can be advanced at a rate proportional the video frame
rate, here $\nicefrac{1}{3}$ of the video frame length, as shown in
Figure~\ref{fg:syncfeat}. This provides an alternate means of getting
synchronized audio-visual features without changing the visual features.
In Section~\ref{frame_rate}, we illustrate the difference in performance
between extracting features at the proposed variable frame rate
vs. fixed frame rate.

\subsection{RNN-T for Audio-Visual Speech Recognition}
\label{sec:rnntav}
\negsectionspace

\begin{figure}[th]
\negfigspace
  \centering
  \includegraphics[width=8cm]{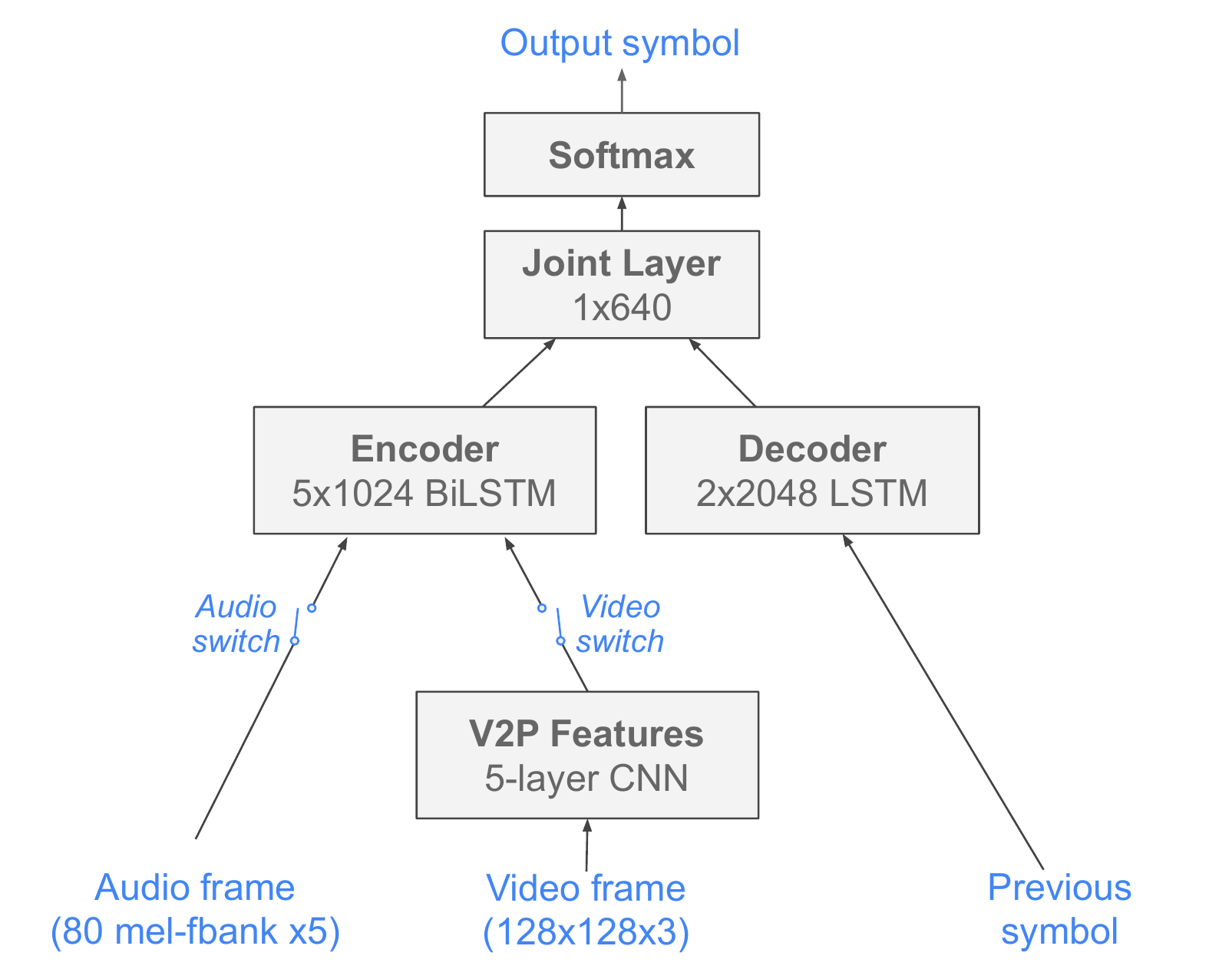}
  \caption{RNN-T model architecture. Isolating modalities is made possible with
  the video and audio switches.}
  \label{fg:rnnt}
\negfigspace
\vspace{-6pt}
\end{figure}

Audio features are derived as discussed in the previous
section~\ref{sec:syncav}. In the audio-only model, the 400 dimensional
stack of filterbank coefficients is fed into the encoder part of the
RNN-T model by turning on only the `audio switch' in
Figure~\ref{fg:rnnt}. In our experiments we use a 5-layer stack of
bidirectional LSTMs in the encoder part of the model with 512 nodes used in each
direction for a total of 1024 nodes in each layer. The decoder is comprised of
2 layers of unidirectional LSTMs each project down to 640 nodes. The joint part
of the RNN-T model combines the encoder and decoder in a 640 dimensional space.
The graphemic output space is 75 characters.

The video model is composed of 5 blocks of $3\times 3 \times 3$
convolutional filters as the features used in
V2P~\cite{dm_lsvsr_2019}.  The number of filters increases from 64 in
the first block, to 128, 256, 512 and 512 in the final layer. After
each convolutional layer, before pooling, group
norm~\cite{groupnorm18eccv} is applied with 32 groups, over the
spatial dimensions, but not the temporal, which gives a small
improvement. The output of the video model yields an embedding vector
of 512 coefficients for every input frame. The video model has 11.7M
parameters. A detailed description of the video model can be found
in~\cite{dm_lsvsr_2018}. To create a visual transcription model, the
512 dimensional output is fed into the encoder by turning on the
`video switch' in Figure~\ref{fg:rnnt}.

For an A/V model, the audio and visual features are enabled
and concatenated to form a 912-dimensional feature vector that is fed
into the encoder as shown in Figure~\ref{fg:rnnt}. A summary of the
parameters, allocated to the various parts of the network for the full A/V
model, is shown in Table~\ref{tb:avparams} minus any temporary
variables used only in training, e.g. moment estimates.  For
simplicity, the shape information listed is for the main kernel of the
layer only and for example does not list the gamma and beta terms from
layer norm, however the parameter count includes them. Layer
normalization~\cite{layernorm16} is used in all the LSTM layers to
make training more stable.

\begin{table}[th]
\negfigspace
  \footnotesize
  \centering
  \quad \\
  \begin{tabular}{llr}
    \toprule
Name            & Kernel Shape & \# Parameters \\
    \midrule
video/block0 & $3\!\times\! 3\!\times\! 3\!\times\! 3\!\times\! 64$ & 5.4K \\
video/block1 & $3\!\times\! 3\!\times\! 3\!\times\! 64\!\times\! 128$ & 221.6K \\
video/block2 & $3\!\times\! 3\!\times\! 3\!\times\! 128\!\times\! 256$ & 885.5K \\
video/block3 & $3\!\times\! 3\!\times\! 3\!\times\! 256\!\times\! 512$ & 3.5M \\
video/block4 & $3\!\times\! 3\!\times\! 3\!\times\! 512\!\times\! 512$ & 7.1M \\
encoder/rnn0 & $1424\!\times\!512\!\times\!4\!\times\!2$ & 5.8M \\
encoder/rnn1 & $1536\!\times\!512\!\times\!4\!\times\!2$ & 6.3M \\
encoder/rnn2 & $1536\!\times\!512\!\times\!4\!\times\!2$ & 6.3M \\
encoder/rnn3 & $1536\!\times\!512\!\times\!4\!\times\!2$ & 6.3M \\
encoder/rnn4 & $1536\!\times\!512\!\times\!4\!\times\!2$ & 6.3M \\
decoder/rnn0 & $715\!\times\!2048\!\times\!4 \!+\! 2048\!\times\!640\!$ & 7.2M \\
decoder/rnn1 & $1280\!\times\!2048\!\times\!4 \!+\! 2048\!\times\!640$ & 11.8M \\
rnnt/encoder & $1024\!\times\!640$& 655.4K \\
rnnt/decoder & $640\!\times\!640$& 409.6K \\
rnnt/output  & $640\!\times\!75$& 48.1K \\
\midrule
Total           &  & 62.9M \\
\bottomrule
\end{tabular}
 \caption{Audio-visual RNN-T model architecture.}
\label{tb:avparams}
\negfigspace
\vspace{-6pt}
\end{table}

\negsectionspace
\section{Experiments and Results}
\label{Experiments}
\negsectionspace

To train the models in this work, we use TensorFlow to implement our
neural network models. An Adam optimizer~\cite{kingma15adam} is used
for training. Unless otherwise noted, we used a learning rate schedule
warming up linearly to $2\times 10^{-3}$ in 20k steps, staying
constant for 50k steps, before exponentially decaying. We train our
models on tensor processing units (TPUs) with 128 cores and a batch
size of 16 sent to each core for an effective batch size per step of
2048 utterances. For all results, we select the model checkpoint that
produces the best result on a held-out validation shard of the
training data. We present results first on a selection of YouTube
utterances where we found they had visual tracks that covered most of
the speech in the segment and were well synchronized. This set of
utterances is collected from over 1000 videos totaling over 20k
utterances and called \ytdev. We then present results on the publicly
available \lrsted test set which contains about 10k words and 1436
utterances.
The results are measured in word error rates (WER) of the output of frame-synchronized beam decoder \cite{prabhavalkar17comparison},
where at most 4 highest scoring candidates are retained at every step during decoding.
CI denotes the half width of the 95\% confidence intervals
based on~\cite{vilar08confidenceinterval}.

\negsectionspace
\subsection{Baseline Results}
\label{sec:baseline}
\negsectionspace
Training an audio-visual model can be difficult since the speech
signal is far more informative than the visual. However, our results
in Table~\ref{tb:yt_baseline_results} show that an A/V RNN-T model can
be trained from scratch on A/V data without pre-training any
components. On the 31k training set, an audio-only sytem can be
trained that yields a 21.5\% word error rate (WER) on \ytdev. In comparison,
the video-only system gives a 48.5\% error rate. In combination, an
A/V system trained on both modalities, leads to a better system at
20.5\%.

\begin{table}[h]
\negfigspace
  \centering
    \begin{tabular}{llc}
      \toprule
      Eval Set & Train \& Eval Mode & WER (\%) $\pm$ 95\% CI \\
      \midrule
              & Audio       & 21.5 $\pm$ 0.5 \\
      \ytdev  & Visual       & 48.5 $\pm$ 0.6 \\
              & Audio+Visual     & 20.5 $\pm$ 0.5 \\
      \bottomrule
    \end{tabular}
    \caption{Results on \ytdev where the training
      modalities match the test.}
   \label{tb:yt_baseline_results}
\negfigspace
\end{table}

\negsectionspace
\subsection{Training with Drop-out}
\label{sec:dropout}
\negsectionspace
Our best results so far have been with an A/V model trained on both
audio and visual data. However it is interesting to see how such
system behaves when one of these modes are missing.  We evaluated our
ASR system that trained from both audio and visual modalities with
unimodal test data. As shown in the first row of results in
Table~\ref{tb:yt_dropout_results}, when the A/V trained model is
evaluated on a audio-only version of the test set, we found that the
performance degrades from 20.5\% to 24.0\% demonstrating that the
model has learned to depend on both modes for recognition. More
surprising, is that when evaluated on a visual-only version of the
test set, the results become extremely poor at 98.8\%. In real-world
data, both modalities may not always be present so it may be prudent
to improve results on unimodal data for robustness.

To boost results with missing modalities, we simulate the effect
during training by randomly dropping out a modality at the sentence
level as in~\cite{zhang19avdropout}. As shown in Table~\ref{tb:yt_dropout_results}, the first row of
results indicates when no drop-out is applied. In the second row,
there is a 30\% chance the audio is dropped during training. This
makes the model learn to use the video signal almost as well as the
visual-only system: 50.3\% compared to 48.5\% and the overall A/V
system is better at 19.8\%. The cost is the results on only audio-only
data are much worse, now at 46\%. To improve the audio-only results,
we also tried training with a 30\% chance of audio being dropped, 10\%
chance of video being dropped, or neither. For this model with audio
and video drop-out, we use a different learning rate schedule
($4\times 10^{-3}$ and kept constant until $100k$ steps) to compensate
for slower learning. This did not have an appreciable affect on the
audio-only and A/V results but surprisingly did harm the visual-only
results by 10\% relative. Thus, for the rest of paper we report
results with 30\% audio dropout, but no visual drop-out. Overall,
these results suggest that it was easier for the network to learn from
audio modality than from visual modality, unless it is forced by
drop-out during training. They also indicate that the current strategy
of modality fusion through the concatenation of features is not robust
to these different input conditions.

\begin{table}[tb]
\negfigspace
  \centering
  \small
  \begin{tabular}{lcc|ccc}
    \toprule
             \multicolumn{3}{r}{Drop-out Chance\!\!\!\!\!\!}
                 & \multicolumn{3}{c}{Eval Mode} \\
    Eval Set & A & V & A & V & A+V \\
    \midrule
               & \!---\!  & \!---\!  & \!24.0 {\footnotesize $\pm$0.5}\! & \!98.8 {\footnotesize $\pm$0.1}\! & \!20.5 {\footnotesize $\pm$0.5}\! \\
    \ytdev     & \!30\%\! & \!---\!  & \!46.0 {\footnotesize $\pm$0.9}\! & \!50.3 {\footnotesize $\pm$0.7}\! & \!19.8 {\footnotesize $\pm$0.5}\! \\
               & \!30\%\! & \!10\%\! & \!45.6 {\footnotesize $\pm$0.9}\! & \!55.3 {\footnotesize $\pm$0.7}\! & \!20.0 {\footnotesize $\pm$0.5}\! \\
    \bottomrule
  \end{tabular}
  \caption{Results with A+V models trained with different drop-out
    rates and tested on audio-only (A), visual-only (V) and
    audio-visual (A+V) modes. All values are WER (\%) $\pm$ 95\% CI.}
  \label{tb:yt_dropout_results}
\negfigspace
\end{table}

\negsectionspace
\subsection{Variable Versus Fixed Frame Rate}
\label{frame_rate}
\negsectionspace

We compared the performance with variable frame rate to those of the
fixed frame rates. For fixed frame rates, we chose 33$\nicefrac{1}{3}$
frames-per-second, to match a conventional 10ms$\times$3 audio frame
system; 30 fps since about half of the videos are 30 or 29.97 fps; and
24 fps to avoid frame duplication.

\begin{table}[h]
\negfigspace
  \centering
  \small
  \begin{tabular}{cc|cccc}
    \toprule
    \multicolumn{2}{c|}{Modes used} & \multicolumn{4}{c}{Frame rate (fps)} \\
    \!Train\!  & \!Eval\!  & variable & 33$\nicefrac{1}{3}$ & 30 & 24 \\
    \midrule
     V     & V     & 48.5 {\footnotesize $\pm$0.6} & 48.6 {\footnotesize $\pm$0.6} & 48.5 {\footnotesize $\pm$0.6} & 50.1 {\footnotesize $\pm$0.6} \\
     \!A+V\!   & \!A+V\!   & 19.8 {\footnotesize $\pm$0.5} & 20.4 {\footnotesize $\pm$0.5} & 20.2 {\footnotesize $\pm$0.5} & 20.1 {\footnotesize $\pm$0.5} \\
     A     & A                & 21.5 {\footnotesize $\pm$0.5}  & 21.6 {\footnotesize $\pm$0.5}  & 21.2 {\footnotesize $\pm$0.5}  & 21.4  {\footnotesize $\pm$0.5} \\
    \bottomrule
  \end{tabular}
  \caption{Results on \ytdev with different frame rate
    conditions.  All values are WER (\%) $\pm$ 95\% CI. }
  \label{tb:frame_results}
\negfigspace
\end{table}

Table~\ref{tb:frame_results} shows the results with different frame
rate conditions.  The difference between these frame rate conditions
are not significant, as with a different learning rate schedule we
observed that fixed 30fps model achieved slightly better accuracy than
variable frame rate model.  We can conclude, using a variable frame
rate is as effective as using fixed frame rates with an RNN-T model\@.

\negsectionspace
\subsection{Impact of Visual Features on Noisy Speech}
\label{sec:noisy_results}
\negsectionspace

\begin{table}[!b]

\negfigspace
  \centering
  \small
  \begin{tabular}{llccc}
    \toprule
      Test set & Added noise         &  A   & A+V &  V \\
    \midrule
                   & ---                   & \!21.5 {\footnotesize $\pm$0.5}\! & \!20.5 {\footnotesize $\pm$0.5}\! &  \\
                   & babble, 20dB               & \!22.5 {\footnotesize $\pm$0.5}\! & \!21.2 {\footnotesize $\pm$0.5}\! &  \\
    \!\!\ytdev\!\!\!
                   & babble, 10dB               & \!28.1 {\footnotesize $\pm$0.5}\! & \!24.8 {\footnotesize $\pm$0.5}\! &  \!\!48.5 {\footnotesize $\pm$0.6}\!\!\\
                   & babble, 0dB                & \!64.5 {\footnotesize $\pm$0.5}\! & \!57.4 {\footnotesize $\pm$0.6}\! &  \\
                   & overlapping speech\!\!\!\! & \!40.6 {\footnotesize $\pm$0.6}\! & \!37.4 {\footnotesize $\pm$0.6}\! &  \\
    \bottomrule
  \end{tabular}
  \caption{Results on noisy versions of \ytdev for audio (A), visual
    (V) and audio-visual (A+V) models trained on uncorrupted training
    data. All values are WER(\%) $\pm$ 95\% CI.}
  \label{tb:yt_results_noisy}
  \centering
  \quad \\
  \begin{tabular}{llccc}
    \toprule
      Test set & Added noise         &  A   & A+V &  V \\
    \midrule
                   & ---                   & \!21.0 {\footnotesize $\pm$0.5}\! & \!20.0 {\footnotesize $\pm$0.5}\! &  \\ 
                   & babble, 20dB               & \!21.9 {\footnotesize $\pm$0.5}\! & \!20.5 {\footnotesize $\pm$0.5}\! &  \\
    \!\!\ytdev\!\!\!
                   & babble, 10dB               & \!26.6 {\footnotesize $\pm$0.5}\! & \!23.1 {\footnotesize $\pm$0.5}\! &  \!\!48.5 {\footnotesize $\pm$0.6}\!\!\\
                   & babble, 0dB                & \!57.7 {\footnotesize $\pm$0.5}\! & \!42.3 {\footnotesize $\pm$0.6}\! &  \\
                   & overlapping speech\!\!\!\! & \!31.5 {\footnotesize $\pm$0.6}\! & \!25.6 {\footnotesize $\pm$0.6}\! &  \\
    \bottomrule
  \end{tabular}
\caption{Results on noisy versions of \ytdev for audio (A), visual
    (V) and audio-visual (A+V) trained models where overlapping speech
    is added to the training audio 10\% of the time. All values are
    WER(\%) $\pm$ 95\% CI..}
  \label{tb:yt_results_train_overlap}
\end{table}
Table~\ref{tb:yt_results_noisy} presents results on noisy datasets. In
all cases, the models are tested with the same set of modalities as
used in training. We find that there is about a 5\% relative
improvement from having an audio-visual model over an audio-only on
the \ytdev test set without added noise. We can see that the advantage
of the multimodal model is greater on the noisier versions of the test
set. The visual-only model performs the same regardless of added
noise, which in the case of the severe 0dB babble noise conditions, is
better than either using audio-only or audio-visual systems. This is
not entirely unexpected because there is no added noise in our audio
training data.

We can artificially add overlapping speech to the audio training data
by randomly selecting an utterance from the training set as a form of
multi-style training~\cite{lippmann87multistyle}. We do so with a 10\%
chance at a level between 0-20dB randomly selected over a uniform
distribution. In Table~\ref{tb:yt_results_train_overlap}, we report
results with models trained in such a manner. The audio-only model
shows improvement from the overlap training across all the noise
condidtions. The audio-visual model shows even greater relative improvements,
especially on the 0db babble and overlapping speech tests. We believe that
providing the visual signal allows for the A/V model to
better ignore competing speech, 
similar to speaker seperation/speech enhancement with visual
features~\cite{l2l18acm,afouras18avenhance}.

\negsectionspace
\subsection{Performance on \lrsted Dataset}
\label{ted_results}
\negsectionspace
To provide a reference on how our models work on a publicly available
data set, we report results on the \lrsted dataset, using our models
trained only on the YouTube 31khr data set, in
Table~\ref{tb:ted_results} along with some previously published
results. The CTC-V2P system in~\citet{dm_lsvsr_2018} was shown to
yield better transcription results than professional human lipreaders
on YouTube video and despite not training on the available training
data had a word error rate of 55.1\% on this test set. Another
comparison is the TM-seq2seq model
from~\citet{ox_davsr_2018}. TM-seq2seq was shown to better than CTC
and demonstrated how combining audio and video could yield improved
results on a large-vocabulary task. Our system, using far more data
and a RNN-T model makes significant improvement over both of these
systems, achieving a 33.6\% word error rate training for video-only
training and testing, and 4.5\% for a combined A/V system. The
improvement in using both audio and visual modalities over audio holds
for this test set as well.

\begin{table}[h]
  \centering
  \begin{tabular}{llcc}
    \toprule
    Training Data    & Model           & Mode   & WER(\%) \\
    \midrule
    YT, 4khrs        & CTC-V2P           & V & 55.1 \\
    \midrule
    BBC+TED, 1.4khrs & TM-seq2seq    & V   & 58.9  \\
    BBC+TED, 1.4khrs & TM-seq2seq    & A   & \phantom{0}8.3  \\
    BBC+TED, 1.4khrs & TM-seq2seq    & A+V & \phantom{0}7.2  \\
    \midrule
    YT, 31khrs & RNN-T              & V    & 33.6 \\
    YT, 31khrs & RNN-T              & A    & \phantom{0}4.8 \\
    YT, 31khrs & RNN-T              & A+V  & \phantom{0}4.5 \\
    \bottomrule
  \end{tabular}
  \caption{Results of the RNN-T model,
    CTC-V2P~\citet{dm_lsvsr_2018} (trained on $\pm30^{\circ}$ face
    rotations) and TM-seq2seq~\citet{ox_davsr_2018} on \lrsted.}
  \label{tb:ted_results}
  \negfigspace
  \vspace{-6pt}
\end{table}

\negsectionspace
\section{Audio-Visual ASR \& AI Principles}
\label{principles}
\negsectionspace
The development of Audio-Visual ASR technology, and more specifically
lip-reading, raises issues related to privacy, especially in light of
the performance of our system on the \lrsted dataset when using the
visual modality only. It should be noted that this level of
performance relates to the nature of the \lrsted task which involves
professionally produced content under good lighting conditions (no
shadows), high video frame-rate ($\geq$ 24 fps), and a cooperative
speaker facing the camera. This is illustrated by the significantly
higher visual-only WER on the \ytdev set (48.5\%), which reflects
``in-the-wild'' conditions, compared to the \lrsted set (33.6\%), which
corresponds to studio-quality recordings.

We also evaluated the performance of our video-only system as a
function of the image quality by labeling utterances from the \ytdev
set as ``high quality'' when the face had a minimum eye distance of 80
pixels, a bounding box diagonal of 300 pixels, a maximum absolute pan
angle of 30 degrees and a tilt angle less than 10 degrees. Everything
else was labeled as ``low quality'', with 70\% of the utterances
falling in that category. The visual-only WER on the low quality
utterances was 57.0\%, to be compared with 37.1\% on the high quality
utterances. We would expect an even greater degradation in performance
on video streams originating from low resolution and low frame rate
devices such as CCTV cameras.

\ifx\blindreview
We are aware of the risk-benefit trade-off for use of this technology
and our work abides by Google AI Principles~\citet{google:ai}, which
define our commitment to develop technology responsibly.
\else
We are aware of the risk-benefit trade-off for use of this technology
and our work abides by Google AI Principles~\citet{google:ai}.
\fi
We are hoping
that this work, by improving the robustness of speech recognition
systems, will increase the reach of ASR technology to a larger
population of users, as well as the development of assistive
technology. One area of interest is enabling people with impaired
speech, such as for example people suffering from Lou Gehrig's
disease, to continue operating speech-enabled devices and equipment
by relying more on the visual modality.

Last, it should be noted that the data and models developed in this
work are restricted to a small group of researchers working on this
project and are handled in compliance with the European Union General
Data Protection Regulation~\citet{gdpr}.

\negsectionspace
\section{Conclusions}
\label{Conclusions}
\negsectionspace
In this paper, we presented an RNN-T based speech recognition system
operating on audio-visual content. We adopted a fully automated
approach to mine a large audio-visual training set out of YouTube
public videos, relying on advances in ASR and computer vision to
select utterances with an on-screen speaking face, its corresponding
audio signal and its matching user-uploaded captions. We used a V2P
frontend to extract visual features from thumbnails located on the
speaker mouth region and enforced extracting audio features at the
same frame rate as the visual features. This enabled concatenating the
audio and visual features at the input of the RNN-T encoder, leading
to a similar modeling architecture for our audio, visual, and AV
system.  We illustrated that our AV system slightly improves over an
audio-only system when trained on the same amount of training data,
but leads to significant performance improvement in presence of babble
noise or overlapping speech. We also described how the use of such a
large training set leads to state-of-the-art performance on the
publicly available \lrsted set. Future work will focus on comparing
the performance of our system with other alternative approaches such
as attention-based and transformed-based models and in exploring the
use of modality-imbalanced training sets.


\label{endofmaintext}

\bibliographystyle{IEEEbib}
\bibliography{av}
\end{document}